# Performance of Bursty World Wide Web (WWW) Sources over ABR [1]

Bobby Vandalore, Shivkumar Kalyanaraman, Raj Jain, Rohit Goyal, Sonia Fahmy
The Ohio State University, Department of CIS
Columbus, OH 43210-1277
Phone: 614-688-4482, Fax: 614-292-2911
Email: {*vandalor, shivkuma, jain, goyal, fahmy*}@cis.ohio-state.edu
Seong-Cheol Kim
Principal Engineer, Network Research Group, Communication Systems R&D Center
Samsung Electronics Co. Ltd., Chung-Ang Newspaper Bldg.
8-2, Karak-Dong, Songpa-Ku, Seoul, Korea 138-160
Email: kimsc@metro.telecom.samsung.co.kr

**Abstract**

We model World Wide Web (WWW) servers and clients running over an ATM network using the ABR (available bit rate) service. The WWW servers are modeled using a variant of the SPECweb96 [1] benchmark, while the WWW clients are based on a model by Mah [2]. The traffic generated by this application is typically bursty, i.e., it has active and idle periods in transmission. A timeout occurs after given amount of idle period. During idle period the underlying TCP congestion windows remain open until a timeout expires. These open windows may be used to send data in a burst when the application becomes active again. This raises the possibility of large switch queues if the source rates are not controlled by ABR. We study this problem and show that ABR scales well with a large number of bursty TCP sources in the system.

## 1 Introduction

As large ATM networks are built, it is important to study the performance of real-world applications like the World Wide Web (WWW) over ATM. Such applications typically have low average bandwidth demands from the network, but care for the response time when active. It is interesting from the traffic management perspective to study the aggregate effect of hundreds of such applications browsing or downloading large documents over an ATM backbone.

The WWW application sets up TCP (Transport Control Protocol) connections for its data transfers [4]. The WWW application differs from a large file transfer application in that while the latter looks like an *infinite* or *persistent* application to TCP, the former looks like a *bursty* application with active and idle transmission periods. The effect of this on traffic management is described below.

TCP increases its *congestion window* as it receives acknowledgements for segments correctly received by the destination. If the application such as file transfer or WWW server/client has data to send, it transmits the data. Otherwise, the window remains open until either the application has data to send or TCP times out, using a timer set by its round trip time (RTT) estimation algorithm. The timer can go off for two reasons (i) retransmission timeout: if an acknowledgement is not received within the timeout period (ii) idle timeout: if there is no activity at the source for a given amount of time. The timeout period is atleast as much as the timer granularity, which is typically 100-500 milliseconds. If the timer goes off off, TCP reduces the congestion window to one segment, and rises exponentially (*slow start* phase) once the source becomes active again.

---

[1]Submitted to the WebNet'97, November 1997. Available through http://www.cis.ohio-state.edu/~jain/papers.html

On the other hand, if the application remains idle for a period smaller than the timeout, the window is still open when the source becomes active again. If acknowledgements (corresponding to the data sent) are received within this idle interval, the window size increases further. Since no new data is sent during the idle interval, the usable window size is larger. The effect is felt when the application sends data in the new burst.

When TCP carrying such a WWW application runs over ATM, the burst of data is simply transferred to the NIC (network interface card). Assuming that each TCP connection is carried over a separate ABR VC (virtual circuit), the data burst is sent into the ATM network at the VC's ACR (allowed cell rate). Since this VC has been idle for a period shorter than the TCP timeout (typically 500 ms for ATM LANs and WANs), it is an *ACR retaining* VC. Source End System (SES) Rule 5 [5] specifies that the ACR of such a VC be reduced to ICR (initial cell rate) if the idle period is greater than parameter ADTF (ACR decrease time-out factor), which defaults to 500 ms. With this default value of ADTF, and the behavior of the TCP application, we are in a situation where the ACR is not reduced to ICR. This situation can be potentially harmful to the switches if ACRs are high and sources simultaneously send data after their idle periods.

Observe that an infinite application using TCP over ABR does not send data in such sudden bursts. As discussed in our previous work [7], the aggregate TCP load at most doubles every round trip time (since two packets are inserted into the network for every packet transmitted, in the worst case). Bursty TCP applications may cause the aggregate load to more than double in a round trip time.

In this contribution, we show that such worst case scenarios are avoided in practice due to the nature of WWW applications and the ABR closed-loop feedback mechanism. In other words, ABR scales well to support a large number of bursty WWW sources running over TCP.

## 2  The WWW System Model

The WWW uses the Hypertext Transfer Protocol (HTTP), which uses traditional TCP/IP, for communication between WWW client and and WWW servers [3].

Modeling of the WWW traffic is a difficult task since the nature of traffic is changing due to the development of new HTTP standards, new WWW servers, WWW clients, and change in user behavior. In this section, we outline our model and the inherent assumptions.

### 2.1  Implications of the HTTP/1.1 standard

The main difference between the latest version of the HyperText Transfer Protocol, HTTP/1.1 [4], and earlier versions is the use of persistent TCP connections as the default behavior for all HTTP connections. In other words, a new TCP connection is not set up for each HTTP request. The HTTP client and the HTTP server assume that the TCP connection is persistent until a *Close* request is sent in the HTTP Connection header field.

Table 1: Class, Files sizes and Frequency of Access

| Class | File Sizes | Frequency of Access |
|---|---|---|
| Class 0 | 0 – 1KB | 20% |
| Class 1 | 1KB – 10KB | 28% |
| Class 2 | 10KB – 100KB | 40% |
| Class 3 | 100KB – 1MB | 11.2% |
| Class 4 | 1MB – 10MB | 0.8% |

Another important difference between HTTP/1.1 and earlier versions is that the HTTP client can make multiple requests without waiting for responses from the server (called *pipelining*). The earlier models were *closed-loop* in the sense that each request needed a response before the next request could be sent.

## 2.2 WWW Server Model

We model our WWW servers as infinite servers getting file requests from WWW clients. The model is an extension of that specified in SPECweb96 benchmark [1]. The file requests fall into five classes (Class 0 through Class 4) which are shown in table 1.

There are nine discrete sizes in each class (eg: 1 KB, 2 KB, on up to 9 KB, then 10 KB, 20 KB, through 90 KB, etc). We assume that the accesses within a class are assumed to be evenly distributed. The model of discrete sizes in each class is based on the SPECweb96 benchmark [1]. The key differences from the SPEC model are (i) the assumptions of an infinite server, and (ii) a new distribution of file sizes, which allows us to model file sizes larger than those in the SPEC benchmark.

Specifically, the average file size in our distribution is approximately 120 kB, compared to an average file size of about 15 kB in SPECweb96. Our distribution introduces an extra class of file sizes (1 MB - 10 MB) which models the downloading of large software distributions, and offline browsing search results. We justify the weight assignments for the various classes in the next subsection.

## 2.3 WWW Client model

Mah HTTP-model [2] describes an empirical model of WWW clients based on observations in a LAN environment. Specifically, a typical client is observed to make, on the average, four HTTP GET requests for a single document. Multiple requests are needed to fetch inline images, if any. With the introduction of JAVA scripts in web pages, additional accesses maybe required to fetch the scripts. Therefore, we use five as the average number of HTTP GET requests. The caching effects at the clients are ignored.

HTTP/1.1 uses persistent TCP connections and sends multiple requests without waiting for the results of the earlier requests (*open-loop mode*). But, the *open-loop* mode is not used for the first request because the additional requests have to be made based on the results of the first response.

Typically, the first HTTP request from a HTTP client accesses the index page (plain text), which is of size 1 KB or less. Since every fifth request is expected to be an index page access, we assign a weight of 20% (= 1/5) for the file size range of 1 KB or less.

Additional requests may require larger responses, as modeled by our distribution of file sizes at servers, taking into consideration the possibility of larger file sizes in the future.

We also model a time lag between batches of requests (presumably for the same document), which corresponds to the time taken by the user to request a new document, as a constant, 10 seconds. While this may be too short a time for a human user to make decisions, it also weights the possibility of offline browsing where the inter-batch time is much shorter.

We do not attempt to model user behavior across different servers. The main purpose of using this simplistic model is to approximate the small loads offered by individual web connections, and to study the effects of aggregation of such small loads on the network.

# 3   The *K-N Client-Server* Configuration

The *K-N Client-Server* configuration shown in Figure 1 has a single bottleneck link (LINK1) shared by the N Web servers and K clients per server or a total of N × K clients). Each client (eg., a netscape browser) sets up a TCP connection which goes through a ATM VC on a separate Network Interface Card (NIC). The VC goes through switch (SW1) to the NIC corresponding to its server. Every server has a NIC card connected to it. Therefore there are N NICs, one for each of the N servers. The ATM VCs from the clients reach the server through the server NIC and a separate TCP entity. Hence, there are K TCP entities corresponding to the K clients per server. In our simulations, K = 15, and N = 1, 2, 5, 10. All link lengths are 1000 km. The bottleneck link (LINK1) speed is 45.0 Mbps, modeling a T3 link, while the remaining link speeds are 155.52 Mbps.

We define feedback delay as sum of the time required for feedback from the bottleneck switch to reach the source and the delay for the new load to be felt at the switch. In our configuration, the feedback delay is 10 ms, which corresponds to 3680 cells at 155.52 Mbps. The round trip time is 30 ms, which corresponds to 11040 cells.

# 4   TCP and ERICA+ Parameters

We use a TCP maximum segment size (MSS) of 512 bytes. The window scaling option is used to obtain larger window sizes for our simulations. For our WAN simulations we used a window of 16 × 64 kB or 1024 kB which is greater than the product of the round trip time (RTT) and the bandwidth yielding a result of 454,875 bytes at 121.3 Mbps TCP payload rate (defined below) when the RTT is 30 ms.

TCP data is encapsulated over ATM as follows. First, a set of headers and trailers are added to every TCP segment. We have 20 bytes of TCP header, 20 bytes of IP header, 8 bytes for the RFC1577 LLC/SNAP encapsulation, and 8 bytes of AAL5 information, a total of 56 bytes. Hence, every MSS of

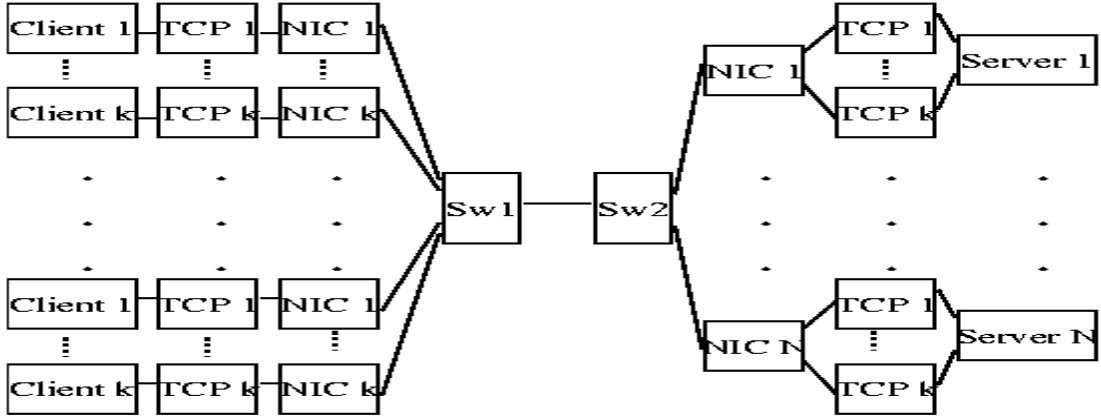

Figure 1: K-N Client-Server configuration

512 bytes becomes 568 bytes of payload for transmission over ATM. This payload with padding requires 12 ATM cells of 48 data bytes each. The maximum throughput of TCP over raw ATM is (512 bytes/(12 cells × 53 bytes/cell)) = 80.5%. Further in ABR, we send FRM cells once every Nrm (32) cells. Hence, the maximum throughput is 31/32 × 0.805 = 78% of ABR capacity. For example, when the ABR capacity is 45 Mbps, the maximum TCP payload rate is 35.1 Mbps. Note that higher efficiency can be achieved by using larger MSS.

We are interested in comparative efficiency, so we normalize our efficiency measurements. We use a metric called *efficiency* which is defined as the ratio of the TCP throughput achieved to the maximum throughput possible. As defined above the maximum throughput possible is 0.78 × (mean ABR capacity).

In our simulations, we have not used the *fast retransmit and recovery* algorithms. Since there is no loss, these algorithms are not exercised.

The ERICA+ algorithm [8] uses five parameters. The algorithm measures the load and number of active sources over successive averaging intervals and tries to achieve 100% utilization with queueing delay equal to a target value. The averaging intervals end either after the specified length or after a specified number of cells have been received, whichever happens first. In our simulations, these values default to 500 ABR input cells or 5 ms. The other parameters are used to define a function which scales the ABR capacity in order to achieve the desired goals. These include a target queueing delay (T0, set to 500 microseconds), two curve parameters (a = 1.15 and b = 1.05), and a factor which limits the amount of ABR capacity allocated to drain the queues (queue drain limit factor, QDLF = 0.5).

# 5 Simulation Results

In our simulations, we use have 15 clients per server (K=15). The number of servers in the system is varied from one to fifteen. Given the average file size of 120 kB, five requests per batch, and a constant inter-batch time of 10 seconds, the average load generated per client is 0.48 Mbps. With N servers (and K=15 clients per server), the expected load on the system is 7.2 × N Mbps. As we vary N from 1 to 15, the expected load increases from 7.2 Mbps to 72 Mbps over a bottleneck link of speed 45 Mbps.

The simulation results are presented in Table 2.

Table 2: Effect Number of Servers on Efficiency and Maximum Queue length

| Sources | ABR Metrics | | |
|---|---|---|---|
| Number of servers | Max Switch Q (cells) | Total TCP Throughput | Efficiency ( % of Max throughput) |
| 1 | 3677 (0.33×RTT Delay) | 6.11 Mbps | 17.4% |
| 2 | 6154 (0.56×RTT Delay) | 14.16 Mbps | 40.3% |
| 3 | 7533 (0.68×RTT Delay) | 21.11 Mbps | 60.1% |
| 4 | 10217 (0.93×RTT Delay) | 28.46 Mbps | 81.1% |
| 5 | 14057 (1.27×RTT Delay) | 34.08 Mbps | 97.1% |
| 6 | 16342 (1.48×RTT Delay) | 34.28 Mbps | 97.6% |
| 7 | 13846 (1.25×RTT Delay) | 34.05 Mbps | 97.0% |
| 8 | 19399 (1.76×RTT Delay) | 34.05 Mbps | 97.0% |
| 9 | 23471 (2.13×RTT Delay) | 33.04 Mbps | 94.1% |
| 10 | 17269 (1.56×RTT Delay) | 32.72 Mbps | 93.2% |
| 11 | 17134 (1.55×RTT Delay) | 33.02 Mbps | 94.1% |
| 12 | 17956 (1.63×RTT Delay) | 32.60 Mbps | 92.9% |
| 13 | 27011 (2.45×RTT Delay) | 32.45 Mbps | 92.4% |
| 14 | 17652 (1.60×RTT Delay) | 30.03 Mbps | 85.6% |
| 15 | 16268 (1.47×RTT Delay) | 30.10 Mbps | 85.8% |

Observe that the maximum switch queue (which corresponds to the buffer requirement at the bottleneck switch) initially *increases linearly* as a function of input load (i.e., number of servers; see rows 1 through 5). The efficiency also increases linearly as a function of load in this range. The efficiency value is high (greater than 85%), the variation in efficiency is due to use of random numbers in the simulation.

However, as the average load on the network exceeds the bottleneck link capacity (N = 7,ldots,15), the buffer requirement does not increase correspondingly. In other words, the *maximum queue stabilizes* under overload conditions due to ABR feedback while the efficiency remains high. This is explained as follows.

Initially, when the network is lightly loaded, switches allocate high rates to sources. Due to the bursty nature of the WWW applications, and the use of persistent TCP connections, the sources may dump bursts of cells into the network (seen as queues at the bottleneck switch). This leads to a *linear increase* in maximum queue lengths for low average loads. The maximum queue length is bounded by 3 × RTT delay cells.

However, as the average load on the bottleneck switch increases, the switch allocates lower rates to the contending sources. Now, even if the WWW applications dump cells, they are not admitted into the ATM network since the source rate allocations are low. The low rate allocations limit the sudden load seen by the network under overload conditions with bursty TCP sources.

# 6  Summary

In this paper, we have investigated the performance of ABR when bursty applications like WWW clients and servers using TCP run over ATM networks. The problem with such applications is that they might utilize open TCP congestion windows to dump bursts of data on the underlying ATM network, resulting in bottleneck queues. Though the burst sizes (captured by file size distributions) used by these applications may be large, the average rate per connection is small. Hence, it takes a large number of connections to load the network. Inspite of control mechanisms at TCP layer and ATM layer the average load increases as the number of sources increase. However, as the load on the network increases, the ABR switch algorithm controls the source rates to low values, and restricts the burstiness in load seen by the network. As a result, the maximum queue lengths in the network are bounded and the efficiency remains high. Hence, ABR responds to smooth load increase and scales well in real-life.